\documentclass[%
 reprint,
superscriptaddress,
 aps,
prl,
]{revtex4-2}
\usepackage{graphicx}
\usepackage{dcolumn}
\usepackage{bm}

\usepackage[utf8]{inputenc}
\usepackage[T1]{fontenc}
\usepackage{mathptmx}
\usepackage{amsmath}
\usepackage{amssymb}
\usepackage{etoolbox}
\usepackage[dvipsnames]{xcolor}

\usepackage{lastpage}

\usepackage{dcolumn}
\usepackage{bm}
\usepackage{hyperref}
\hypersetup{
    colorlinks=true,
    linkcolor=RoyalPurple,
    citecolor=blue,
    urlcolor=black
}
\usepackage{soul}
\newcommand\ins[1]{{\color{black}#1}}

\makeatother
\begin{document}

\preprint{APS/123-QED}

\title{Measuring Absolute Velocities from Non-Equilibrium Oscillations via Single-Detector 3D Dynamic Light Scattering}

\author{Jos\'e L\'opez-Molina}

 \affiliation{ 
Department of Applied Physics, Universidad de Granada, Campus Fuentenueva S/N, 18071 Granada, Spain.
}
\author{Arturo Moncho-Jord\'a}
 \email{moncho@ugr.es.}

 \affiliation{ 
Department of Applied Physics, Universidad de Granada, Campus Fuentenueva S/N, 18071 Granada, Spain.
}
\affiliation{Institute Carlos I for Theoretical and Computational Physics, Universidad de Granada, Campus Fuentenueva S/N, 18071 Granada, Spain.}
\author{Mar\'ia Tirado-Miranda}
 \email{mtirado@ugr.es.}
 
 \affiliation{ 
Department of Applied Physics, Universidad de Granada, Campus Fuentenueva S/N, 18071 Granada, Spain.
}

\date{\today}

\begin{abstract}

Single-detector 3D dynamic light scattering (3D-DLS) emerges as a reliable technique to determine the drift velocity of out-of-equilibrium colloidal particles. In particular, our investigation reveals the appearance of oscillations of a well-defined frequency in the autocorrelation function of the scattered intensity when particles are immersed in a medium exposed to thermally induced convection. These oscillations arise as a consequence of the directed motion of particles due to the convective motion of the fluid. The experimental results obtained for different colloidal systems are corroborated by a theoretical model and thoroughly validated with fluid dynamics and Brownian dynamics simulations. The excellent agreement between experimental, theoretical and simulation data allows us to provide a solid and comprehensive explanation of the observed physical phenomena. This study via advanced dynamic light scattering (DLS)
technique offers insights into the field of non-equilibrium particle dynamics, applicable not only to colloidal suspension affected by steady-state diffusion-convection but also to other non-equilibrium situations, such as systems driven by external fields (gravitational, electric or magnetic fields, among others).

\end{abstract}

\maketitle

Dynamic Light Scattering (DLS) is a powerful tool for measuring dynamic properties over a wide range of systems. Recently, this technique has been employed to study acousto-responsive microgels under ultrasonic influence~\cite{stock2023impact}, flow-induced dynamics~\cite{feng2023dynamic}, and 3D imaging of heterogeneous diffusion~\cite{lee2012dynamic}. Similar techniques such as X-ray Photon Correlation Spectroscopy (XPCS) has also been refined to observe ballistic dynamics in magnetic fields~\cite{schavkan2017dynamics}, direction-dependent diffusion~\cite{wagner2013direction}, non-invasive studies in living cells~\cite{chushkin2022probing} or in colloidal palladium.~\cite{thurn1996photon}. So, the techniques that have yielded valuable insights into the dynamic behaviors of mesoscopic systems are now being adapted to more complex scenarios such as colloidal systems out of equilibrium.

A common phenomena in many particle systems exposed to external fields is the emergence of a particle drift velocity.
Traditionally, these velocities have been obtained with the heterodyne method (HeDLS), which builds the intensity autocorrelation function, $g^{(2)}(\textbf{q},\tau)$, by mixing scattered light from particles with an unscattered reference beam \ins{in a detector}. Particles velocity is derived from the time oscillations that appear in $g^{(2)}(\textbf{q},\tau)$.~\cite{pecora1985dynamic,berne2000dynamic, ito2004comparison, medrano2008measuring, savchenko2019determination} Alternatively, in a classical \ins{homodyne setup} (HoDLS), which uses a single incident beam and a single detector to analyze only the scattered light from particles, theory predicts no oscillatory behavior. ~\cite{pecora1985dynamic,berne2000dynamic} However, several studies have observed oscillations using HoDLS, leading a considerable discussion about its physical interpretation. ~\cite{moulin2020homodyne, josefowicz1975homodyne, sehgal1999anomalous} 

\begin{figure}[htbp]
\includegraphics[width=\linewidth]{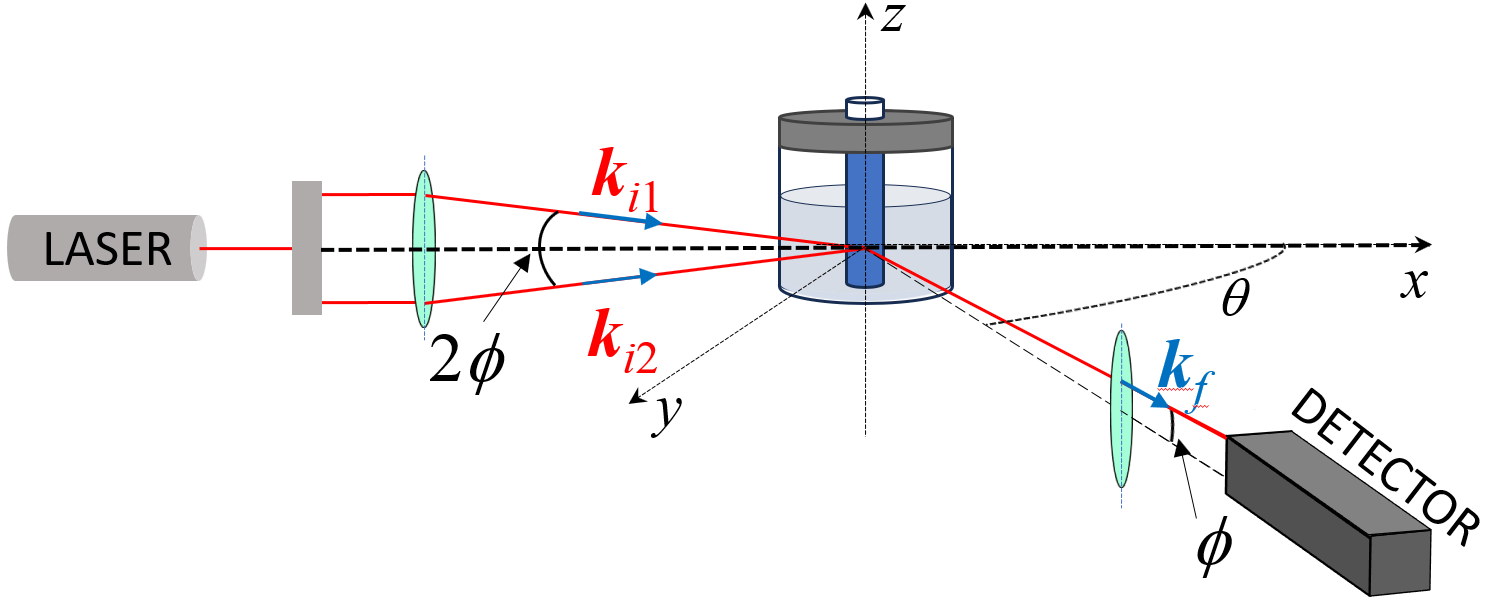}
\caption{Experimental scattering setup. The incident laser beam is split into two beams that converge at the center of a sample cell. The angle between each beam and the scattering plane is denoted as $\phi$. The scattered light is collected by a detector positioned at an angle $\theta$ relative to the initial laser direction and $\phi$ above the plane.
}
\label{fig:dispositivo}
\end{figure}

The 3D-DLS setup is a standard device widely used in many laboratories to characterize dense colloidal suspensions.~\cite{schatzel1991suppression, urban1998characterization} In this work, we expand its conventional application to include the detection of particle drift velocities induced by external fields using a 3D-DLS setup with a single detector which collects the scattered light from two intersecting laser beams within the sample (see Fig.~\ref{fig:dispositivo}). Based on light scattering theory, we provide a formal justification for the observed oscillatory behavior of $g^{(2)}(\textbf{q},\tau$) arising in non-equilibrium steady-state colloidal fluids, and connect this phenomena to the existence of a drift velocity.
This setup enables the precise quantification of drift velocities from the new intensity autocorrelation function. We illustrate this procedure by studying the drift velocity in fluid-induced thermal-convective systems, while emphasizing its applicability to other non-equilibrium scenarios. Our experimental findings and theoretical description are fully validated by Brownian dynamics (BD), and fluid dynamic simulations.

Before presenting the experimental results, we provide the theoretical framework required to understand the behavior of $g^{(2)}(\textbf{q},\tau)$ when the  particles not only move by Brownian diffusion, but also due to an imposed drift velocity. We start with the simplest HoDLS device, namely the one with a single beam and a single detector. The field autocorrelation function of a monodisperse system of spherical particles with diffusion coefficient $D$ and drift velocity $\mathbf{v}$ is given by~\cite{pecora1985dynamic,moulin2020homodyne, sehgal1999anomalous, katayama2009diffusion}
\begin{equation}
    g^{(1)}(\textbf{q},\tau)=e^{i\mathbf{q}\cdot\mathbf{v}}e^{-Dq^2\tau},
    \label{eq:g1_mono}
\end{equation}
where $\mathbf{q}$ is the scattering vector of the monochromatic laser beam. With the traditional one-beam HoDLS it is only possible to measure the intensity correlation function, given by~\cite{dhont1996introduction, goldburg1999dynamic, berne2000dynamic, pecora1985dynamic}

\begin{equation}
    g^{(2)}(\textbf{q},\tau)=1+\beta|g^{(1)}(\textbf{q},\tau)|^2=1+e^{-2Dq^2\tau},
    \label{eq:no_oscila}
\end{equation}
\ins{where $\beta$ is the coherence factor. In the following equations, we set $\beta=1$ for simplicity although in experiments it is a number $0<\beta<1$.} As observed, $\mathbf{v}$ does not explicitly appear in $g^{(2)}$.~\cite{fuller1980measurement, bicout1994multiple} However, in the study of velocity driven systems, a non-equilibrium anomaly is observed in the correlation function that is not captured by Eq.~(\ref{eq:no_oscila}). In particular, the diffusion coefficient derived from the experimental decay of $g^{(2)}(\tau)$ consistently exceeds the value predicted by the Stokes-Einstein equation, $D_{\text{st}} = k_BT / (6\pi \eta a)$ (where $k_B$ is the Boltzmann constant, $T$ the absolute temperature, $\eta$ is the solvent viscosity and $a$ denotes the radius of the colloidal particles). In other words, $g^{(2)}(\tau)$ decays faster with time than expected for this particle size. This phenomenon has been previously attributed to an enhancement of the diffusion coefficient due to reduced friction caused by the convective flow~\cite{katayama2009diffusion} or discussed as superdiffusion.~\cite{carl2019normal} In this work, we show that the rapid decay of $g^{(2)}(\tau)$ is provoked to particles systematically escaping the scattering volume due to the drift velocity, resulting in a loss of correlation.

A proposed correction involves an exponential term dependent on velocity and time squared, based on a Gaussian laser beam intensity profile.~\cite{chowdhury1984application, sun2021nonuniform, leung2006particle, busch2008dynamics, torquato2023microfluidic} However, this approach fails to accurately measure the diffusion coefficient for low drift velocities in our experiments. We reconsidered the laser beam profile's impact, focusing on the scattering volume defined by the intersection of laser beams and the detector's pinhole projection. This led us to treat the beam's intensity as uniform within the scattering region and 0 outside it.

The amplitude of the scattered electric field is thus considered constant across the beam thickness $h$, i.e. $P(z) = E_0$  for $ -\frac{h}{2} < z < \frac{h}{2} $. This revised model introduces a new factor into the correlation function, $\langle P(z(0)) P(z(t)) \rangle g^{(1)}(\mathbf{q},\tau)$, where  $g^{(1)}(\mathbf{q},\tau)$  is given by Eq.~(\ref{eq:g1_mono}) and $z(t) = vt + z_0$, accounting for the observed diffusion coefficient increase by considering the velocity, $v$, effect and $h$. The modified field correlation function for a single-beam device is expressed as:
\begin{equation}
    g^{(1)}(\textbf{q},\tau)=e^{i\textbf{q}\cdot\textbf{v}}e^{-Dq^2\tau}\left(1-\frac{v\tau}{h}\right) \ \ \ \ \  \tau \leq h/v,
    \label{eq:g2_mono_apendice}
\end{equation}
and $g^{(1)}(\textbf{q},\tau) = 0$ if $\tau>h/v$. The term $1-\frac{v\tau}{h}$ embodies the rate at which particles drift out of the scattering volume.

Having studied the single-beam device, we describe our setup with one detector but two beams with scattering vectors $\mathbf{q}_1$ and $\mathbf{q}_2$, respectively, that interfere inside the sample volume (see Fig.~\ref{fig:dispositivo}). Here only the most relevant equations are shown. The full theoretical demonstration is available in the Sec. A of the Supplemental Material (SM). Using that photons scattered by particles from different beams are not intercorrelated, $g^{(2)}$ is generalized to~\cite{schatzel1991suppression, aberle1998effective, phillies1981suppression, pusey1999suppression}
\begin{equation}
    g^{(2)}(\textbf{q}_1,\textbf{q}_2,\tau)-1 = |A_1g^{(1)}(\textbf{q}_1,\tau)+ A_2g^{(1)}(\textbf{q}_2,\tau)|^2,
    \label{eq:g1_doshaces_fin}
\end{equation} 
where $A_1$ and $A_2$ are constants related to each beam intensity and alignment.
Combining Eqs.~(\ref{eq:g2_mono_apendice}) and (\ref{eq:g1_doshaces_fin}), a new expression arises that gathers both the oscillatory behavior and the enhanced loss of correlation for the two-beams arrangement:
\begin{eqnarray}
\label{eq:g2_DLS_experimental_better}
    g^{(2)}(\textbf{q}_1, \textbf{q}_2, \tau)-1=&\\
    =e^{-2D\bar{q}^2\tau}  &\left[(1-C)+C\,\text{cos}(\Delta\textbf{q}\cdot\textbf{v}\tau)\right] \left(1-\frac{v\tau}{h}\right)^2 \nonumber
\end{eqnarray}
for $\tau < h/v$ and $g^{(2)}(\textbf{q}_1, \textbf{q}_2, \tau)-1=0$ for $\tau > h/v$, where $\bar{q}=|\mathbf{q}_1+\mathbf{q}_2|/2$ represents the modulus of the mean of both scattering vectors, $\Delta\textbf{q}=\mathbf{q}_1-\mathbf{q}_2$ is the scattering vectors' difference, and $0 \leq C \leq 0.5$  represents the relative intensity of each beam. Crucially, this model reveals an oscillatory component, with a characteristic frequency given by $\omega=\Delta\textbf{q}\cdot\textbf{v}$.
As $\omega$ depends on $\textbf{v}$ through this dot product, it is only possible to access the projection of the velocity along the $\Delta\textbf{q}$ vector. In our experimental setup $\Delta\textbf{q}$ has only a  $z$-component (i.e. $z$ corresponds to the vertical direction), so this measurement technique will only allow us to obtain $v_z$.

In our experiments, we study the oscillatory behavior and anomalous loss of correlation of $g^{(2)}$ for a colloidal system affected by a thermally-induced convective flow. For this purpose, we use a 3D-DLS setup from LS-Instruments. This device divides an incoming laser beam of wavelenght $\lambda = 632.8\,{\rm nm}$ into two, slightly shifted above and below the scattering plane (see Fig.~\ref{fig:dispositivo}). As mentioned, our method involves using just one detector, enabling it to capture scattered light from both beams that intersect inside the colloidal sample, resulting in two distinct scattering vectors. As colloidal systems, we use monodisperse polystyrene spheres (from microparticles GmbH, Berlin, Germany) and microgel pNIPAM-co-3BA@ particles~\cite{garcia2020magnetically} in water. Measurements took place in a cylindrical glass cell by LS-Instruments, $0.8$~cm in diameter and $7.5$~cm height with the sample filling up to $4$~cm height. To induce thermal convection, only the lower $1$~cm portion of the cell was immersed in a temperature-controlled bath, with the upper part in contact with metal. Measurements are performed at the center of the cell, $0.5$~cm from the bottom, where the beams of the dual-laser setup intersect. Due to the axial symmetry of our cylindrical cell, fluid velocity in the center of the cell has only $z$-component, meaning  $\textbf{v}=v_z\mathbf{\hat{k}}$. In all measurements the duration is $60$~s and the room temperature is $T_\mathrm{Room}=23^\circ$C.

\begin{figure}[htbp]
\includegraphics[width=\linewidth]{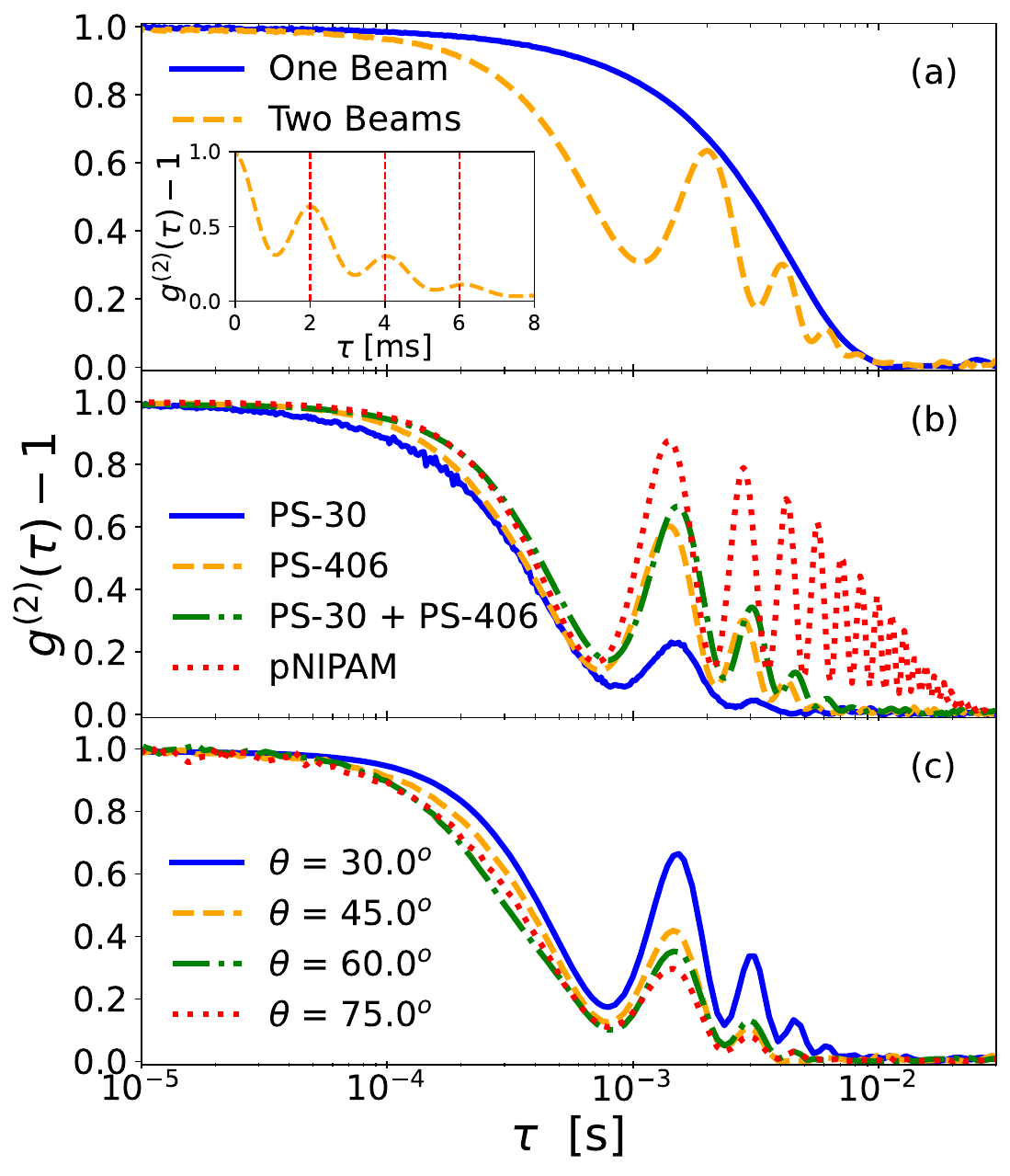}
\caption{Intensity autocorrelation functions, $g^{(2)}(\tau)-1$, obtained via 3D-DLS setup with one detector. Plot (a) shows the effect of the number of beams for $T_\mathrm{Bath} = 38 ^\circ$C and $\theta = 30 ^\circ$. Plot (b) depicts the oscillatory effect for different kind of colloidal particles and sizes, for $T_\mathrm{Bath} = 40 ^\circ$C and $\theta = 30 ^\circ$. Panel (c) shows that the oscillation frequency does not depend on $\theta$, for $T_\mathrm{Bath} = 40 ^\circ$C. In all cases $T_\mathrm{Room}=23^\circ$C.}
\label{fig:big_figure}
\end{figure}

Fig.~\ref{fig:big_figure}(a) presents a comparison between single and dual-beam laser measurements with a bath temperature $T_{\mathrm{Bath}} = 38 ^\circ$C. In the single-beam setup, $g^{(2)}(\tau)$ decays exponentially, as predicted by Eq.~\ref{eq:no_oscila}. In contrast, the dual-beam setup reveals clear oscillations of a well-defined frequency (see inset of Fig.~\ref{fig:big_figure}(a)), confirming that the simultaneous use of two laser beams is essential for observing this phenomenon.

The oscillations observed in our experiments is not a particular result tied to specific particle types; rather, it is a general behavior evident across a range of systems, including $a=406$~nm and $30$~nm radius monodisperse polystyrene particles, a mixture of them, and significantly larger pNIPAM microgels, as illustrated in Fig.~\ref{fig:big_figure}(b).
Particularly notable is the pronounced oscillations in the microgel system, attributable to its larger size. Examining the size dependence of these oscillations reveals that, as particle size decreases, the amplitude of the oscillations dampens. Indeed, $g^{(2)}(\tau)$ decays faster for particles with a larger diffusion coefficient (i.e., particles of smaller size), thus smoothing out the oscillations.

Although convection and oscillations are usually reported as issues related to particle light absorption,~\cite{parashchuk2011hyperdiffusive, moulin2020homodyne, sehgal1999anomalous, schaertl1999convection} we see that thermal gradients inducing convection and oscillations can be externally imposed and measured also in non-light absorbing colloids, regardless of particle composition. We observe this phenomenon in both monodisperse and bidisperse systems, indicating that all particles, regardless of size, move at a common velocity imposed by the convective flow of the solvent. The study by Moulin et al.~\cite{moulin2020homodyne} also reported oscillations in the correlation function of a system consisting of bidisperse light-absorbing particles. These oscillations were attributed to the existence of two distinct velocities for large and small particles. We propose that the cause of these oscillations aligns with our findings: a single drift induced by the convective solvent flow  with two laser beams incident on the system $ \Delta\textbf{q}\cdot \textbf{v}$.
However, if the system exhibits two distinct velocities and only one laser beam is employed, it is plausible that oscillations with a frequency of 
$\textbf{q}\cdot \Delta \textbf{v}$ will emerge, thereby enabling the measurement of the velocity difference between both species (see Sec. B of the SM). This two-velocity scenario may occur in cases involving sedimentation with two types of particles, or if two charged particles move in the presence of an electric field and have different responses to the field.

The influence of the scattering angle, $\theta$, on the autocorrelation function is illustrated in Fig~\ref{fig:big_figure}(c), for a system comprised by polystyrene particles of radius $a=406\,\text{nm}$. As observed, the oscillation amplitude decreases when increasing $\theta$, but the frequency becomes totally unaffected by changes in $\theta$. To elucidate both effects, we calculate $\Delta \mathbf{q} = \mathbf{q}_1 - \mathbf{q}_2$, where $\mathbf{q}$ is the scattering vector defined by $\mathbf{q} = \mathbf{k}_i - \mathbf{k}_f$ and its modulus is $\bar{q} = \frac{4\pi n}{\lambda}\text{sin}\left(\frac{\theta}{2}\right)$. 
The configuration setup of our 3D-DLS device results in $\Delta \mathbf{q}=(2\pi n/\lambda)(0, 0, 2\sin(\phi))$, where $\phi=0.052$~rad. As observed, $\Delta \mathbf{q}$ only has a $z$-component (perpendicular to the scattering plane) and does not depend on $\theta$.  This invariance in frequency with the detection angle aligns with our theoretical model, which posits that $\omega=\Delta \mathbf{q} \cdot \mathbf{v}$. In addition, according to Eq.\ref{eq:g2_DLS_experimental_better} the amplitude of the oscillations is modulated by $\exp{(-2D\bar{q}^2\tau})$, so increasing $\theta$ increases $\bar{q}$, reducing observable oscillations. Therefore, to detect small drift velocities, low angles are preferable.

\begin{figure}[htbp]
\centering
\includegraphics[width=\linewidth]{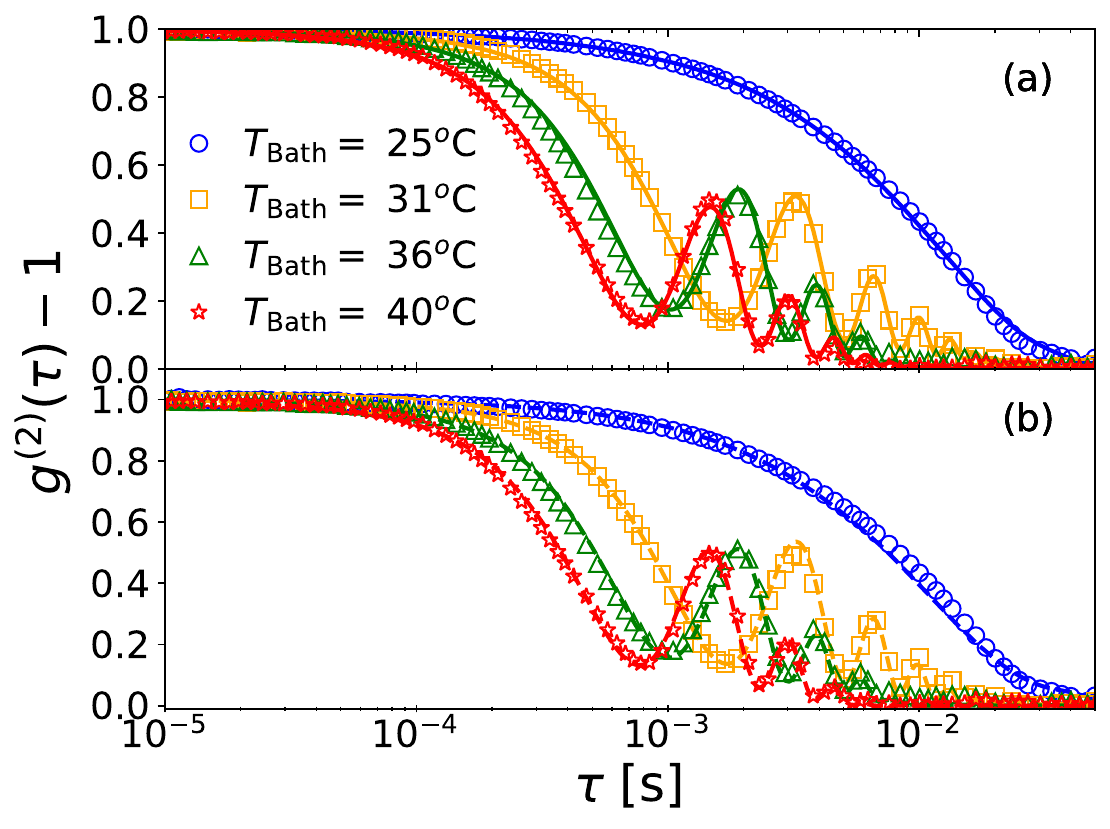}
\caption{$g^{(2)}(\tau)-1$ as a function of $\tau$ for particles with a radius of $a=406$~nm, illuminated by two beams at different $T_{\mathrm{Bath}}$. Symbols represent the experimental measurements obtained with the 3D-DLS device with only one detector. In plot (a), solid lines depict fits of the experimental data to the theoretical model (Eq.~(\ref{eq:g2_DLS_experimental_better})). In plot (b), dashed lines represent correlation functions derived from BD simulations at velocities determined by the respective fits. In both panels $\theta = 40^\circ$, and $T_\mathrm{Room}=23^\circ$C.}
\label{fig:Fit}
\end{figure}

Fig.~\ref{fig:Fit}(a) shows the experimental correlation function (symbols) for different $T_\textmd{Bath}$. When $T_\textmd{Bath}$ is close to $T_\textmd{Room}$, $g^{(2)}$ does not show oscillations. 
This observation indicates that the convective currents are indeed driven by the thermal gradient. As $\Delta T_\textmd{BR}=T_\textmd{Bath}-T_\textmd{Room}$ is increased, we observe the emergence of oscillations, which shift towards the left. This shift signifies an increase in the vertical velocity of the particles, $v_z$.
\begin{table}[htbp]
    \centering
    \begin{tabular}{l|c|c|c|c}
        $T_{\mathrm{Bath}}$ [$^\circ$C] & $D_\textmd{st}$ [{\textmu}m$^2$/s] & $D_\textmd{fit}$ [{\textmu}m$^2$/s] & $v_z$ [mm/s] & $C$\\\hline
         25& 0.596 & 0.56 & 0.07 & 0.41 \\
         31& 0.693 & 0.71 & 1.36 & 0.40\\
         36& 0.780 & 0.86 & 2.30 & 0.36\\
         40& 0.901 & 0.97 & 2.95 & 0.39\\
    \end{tabular}
    \caption{Values of $D_\mathrm{fit}$, $v_z$ and $C$ obtained by fitting the experimental data from Fig.~(\ref{fig:Fit}) using Eq.~(\ref{eq:g2_DLS_experimental_better}) for different $T_{\mathrm{Bath}}$. The corresponding Stokes-Einstein diffusion coefficients, $D_\mathrm{st}$, are also included for comparison.}
    \label{tab:Fitted data}
\end{table}

The experimental results are compared with theoretical prediction as outlined by Eq.~(\ref{eq:g2_DLS_experimental_better}), depicted by solid lines in Fig.\ref{fig:Fit}(a). The fitting procedure is performed as follows. First, we extract the drift velocity from the oscillation using the relation $v_z=\omega/|\Delta \textbf{q}|$, which allows a very accurate determination of $v_z$. Then, this value is fixed to fit the whole autocorrelation function using the diffusion coefficient ($D_\textmd{fit}$) and constant $C$ as fitting parameters, with the beam thickness set at $h = 30$$\mu$m. The resulting values of $v_z$, $D_\textmd{fit}$ and $C$ are presented in Table \ref{tab:Fitted data}. This table further lists the expected Stokes-Einstein diffusion coefficients, $D_\textmd{st}$, calculated for different temperatures. As observed, Eq.~(\ref{eq:g2_DLS_experimental_better}) captures the experimental data in all cases. The fitted $D_\textmd{fit}$ values  closely match the expected $D_\textmd{st}$, with discrepancies ranging from 5\% to 8\%. Crucially, neglecting the correction for particle drifting out the dispersion volume 
\ins{$\left(1-\frac{v\tau}{h}\right)$}
leads to $D_\textmd{fit}$ values inaccurately increasing with temperature $T_{\mathrm{Bath}}$, deviating from $D_\textmd{st}$ values and significantly elevating the error over 100\% at high drift velocities. The velocities obtained align with typical values for such measurement cells.~\cite{ruseva2018capillary} The parameter $C$ was observed to stay consistently around $0.4$, indicating the reliability of the theoretical framework. It was not possible to reproduce velocity measurements for $T_\textmd{Bath}>40^\circ$C, since the system reaches unsteady convection.

To further validate the proposed explanations, we conduct Brownian dynamics simulations on an ideal system with $N=10^4$ colloidal particles, in which particles possess the expected diffusion coefficient, $D_\mathrm{st}$. The simulation employs a time step of $10^{-5}$~s during a total time of $10$~s and is set in a simulation cubic box with side $h =30$~$\mu$m representing the scattering volume.
A vertical upward flow is imposed along the $z$-axis representing the fluid velocity due to thermal convection. The associated values of $v_z$ are extracted from our fits presented in Table~\ref{tab:Fitted data}. Periodic boundary conditions are applied on the $x$ and $y$ directions.  When a particle exits the simulation box through the top, another one is randomly introduced at the bottom, maintaining constant the number of particles. The total scattered electric field of the system at each time step is calculated using $ E_s(t) = \sum_{i=1}^N \left[ C \exp(i \mathbf{q}_1 \cdot \mathbf{r}_i(t)) + (1-C) \exp(i \mathbf{q}_2 \cdot \mathbf{r}_i(t)) \right]$. The correlation function of $ E_s(t) $ is then determined and normalized. To account for our particular experimental setup,  $C = 0.4$ is used in the calculation of the autocorrelation function.

Fig.~\ref{fig:Fit}(b) presents the simulated correlation functions (dashed lines) alongside the experimental data (symbols), showing a remarkable agreement, thus confirming that our simulation accurately captures the loss of correlation arising when particles escape from the scattering volume due to the convective flow. BD simulations also reproduce correctly the oscillation frequency obtained experimentally for all $T_\textmd{Bath}$. This evidence robustly supports the full validation of our model for measuring drift velocities in colloidal systems.

Although our DLS method has been used to characterize the dynamics of colloidal suspensions affected by thermal convection, it can also be applied to other non-equilibrium colloidal systems where a vertical velocity appears, such as colloidal sedimentation or electrophoresis. In addition, it can be used for accessing solvent characteristics including viscosity and relevant dimensionless numbers of the fluid, such as Reynolds, P\'eclet and Rayleigh numbers. Moreover, the proposed dual-beam setup  allows very accurate particle velocity determination through the oscillation frequency, and use it to  characterize the correlation loss due to the escape of particles from the scattering volume, which is something very difficult to perform from the fitting of $g^{(2)}(\tau)$ for small velocities in ordinary light scattering setups.\cite{torquato2023microfluidic}

\begin{figure}[htbp]
\centering
\includegraphics[width=\linewidth]{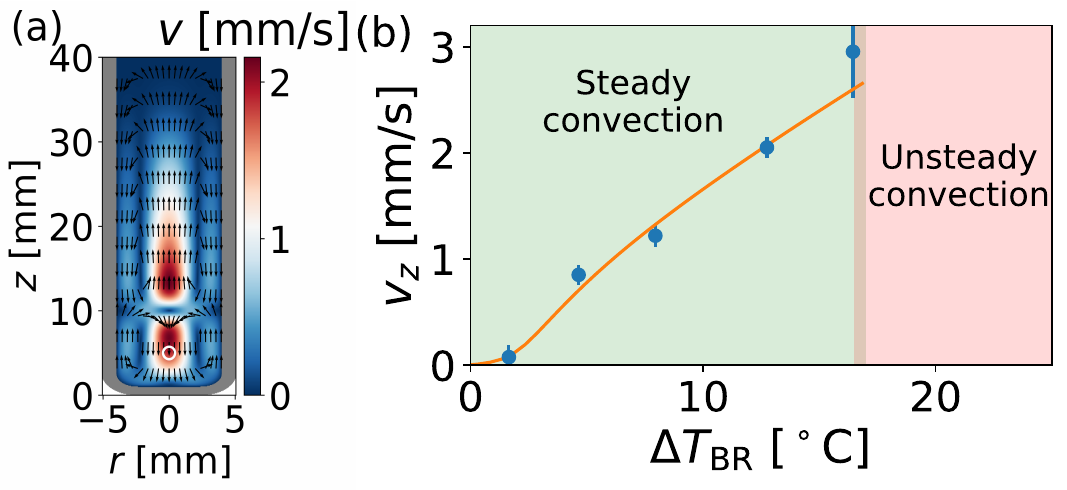}
\caption{Predictions obtained by COMSOL simulations. (a)  Velocity profile in stationary conditions, for a temperature difference given by $\Delta T_\mathrm{BR}=10^\circ$C. The white circle denotes the measurement point in our DLS device. (b) $v_z$ at the measurement point as a function of the temperature difference , $\Delta T_\mathrm{BR}$ (line). Symbols represent the experimental data, obtained from the average over ten independent DLS measurements using Eq.~(\ref{eq:g2_DLS_experimental_better}).}
\label{fig:comsol}
\end{figure}

Using the Nonisothermal Flow interface in COMSOL, free convection was modeled in water. We use this simulation technique to explore the system response over time for different temperature differences, from $\Delta T_\mathrm{BR} =0^\circ$C to $\Delta T_\mathrm{BR} =25^\circ$C. For $\Delta T_\mathrm{BR} \leq 17^\circ$C, fluid dynamics simulations reveal that both the velocity and temperature profiles inside the cell eventually reach a steady state. The stationary velocity field for $\Delta T_\mathrm{BR}=10^\circ$C is illustrated in Fig.~\ref{fig:comsol}(a). We notice two convective regions above and below the bath level that arise as a consequence of the different boundary temperatures, $T_\textmd{Room}$ and $T_\textmd{Bath}$, respectively. The experimental measurement point is highlighted by a circle, inside which magnitudes vary less than 1\%, and the drift velocities are completely vertical. The temperature in this region is lower than $T_\mathrm{Bath}$ due to the convection currents. 

Fig.~\ref{fig:comsol}(b) shows $v_z$ at the measurement point obtained both experimentally and from simulations, as a function of the $\Delta T_\mathrm{BR}$. We clearly identify two distinct dynamic regimes of convective behavior, delineated by the value of $\Delta T_\mathrm{BR}$. In the first regime, corresponding to $\Delta T_\mathrm{BR} \leq 17^\circ$C, the flow eventually reaches a laminar steady state. 
We see excellent quantitative agreement in $v_z$ between simulations and experiments underscoring the robustness of our DLS method for measuring velocities.
For $\Delta T_\mathrm{BR} > 17^\circ$C, the obtained experimental $g^{(2)}(\tau)$ shows huge temporal fluctuations that make impossible to identify the oscillations. Through COM simulations we confirm that, in this dynamical regime, the system keeps on an unsteady state,~\cite{chandrasekhar2013hydrodynamic, clever1974transition, patterson1980unsteady,oresta2007transitional} in which the fluid velocity at the measurement point is not constant anymore and fluctuates over time (see Fig. S2 of SM). This non-steady regime has associated Rayleigh numbers larger than the critical Rayleigh number for this cell geometry (see  Sec. C of the SM).~\cite{le1990transition} Therefore, COMSOL simulations not only reproduce the experimental drift velocities without any fitting parameter, but also capture the transition temperature separating the steady and non-steady dynamic regimes.

In summary, this study extends the conventional applications of 3D-DLS setups using a single detector for measuring drift velocities in non-equilibrium colloidal systems where particles do not only diffuse by Brownian motion, but also posses a drift velocity imposed by thermal or concentration gradients and/or applied external fields.
A notable finding is the elucidation of the underlying physical mechanisms driving the oscillatory behavior of the autocorrelation function. Additionally, our research provides a correction methodology for the loss of correlation observed in this kind of systems via HoDLS, and enriches the understanding of non-equilibrium particle dynamics in such systems, contributing to a broader application of 3D-DLS as a valuable tool in exploring complex fluid dynamics. 

\begin{acknowledgments}
We are grateful to Prof. Antonio M. Puertas-Lopez (University of Almería) and Prof. Alberto Fernández-Nieves for their insightful comments. The authors acknowledge grant PID2022-136540NB-I00 funded by
MICIU/AEI/10.13039/501100011033 and ERDF \textit{A way of making Europe}. J.L.M. thanks the Ph.D. student fellowship (FPU21/03568) supported by the Spanish \textit{Ministerio de Universidades}.
\end{acknowledgments}
    
\bibliography{paper_Lopez_Moncho_Tirado}

\end{document}